\documentstyle[twocolumn,aps,epsfig]{revtex}
\tighten
\begin{document}

\twocolumn[\hsize\textwidth\columnwidth\hsize\csname @twocolumnfalse\endcsname
\title{Prewetting transitions of Ar and Ne on alkali metal surfaces}
\author{Francesco Ancilotto and Flavio Toigo} 
\address{Istituto Nazionale per la Fisica della Materia and 
Dip.to di Fisica G.Galilei \\ Universit\'{a} di Padova, via Marzolo 8, 
35131 Padova, Italy}
\maketitle
\begin{abstract}

We have studied by means of 
Density-Functional calculations the wetting properties of Ar and Ne
adsorbed on a plane whose adsorption
properties simulate the Li and Na surfaces. We use reliable ab-initio
potentials to model the gas-substrate interactions. 
Evidence for prewetting transitions is found for all
the systems investigated and their wetting phase diagrams 
are calculated.

\bgroup\draft
\egroup
\end{abstract}
]

\section{Introduction}

Direct evidence for the first-order nature of the wetting transition, as
originally predicted by Cahn\cite{cahn}, together with the characteristic
prewetting jumps away from the coexistence line, has been obtained in
several different systems: quantum liquid films physisorbed on heavy alkali
metals\cite{taborek}, complex organic liquids\cite{bonn}, near-critical
liquid mercury\cite{fisher} and binary liquid crystal mixtures\cite
{lucht,esteve}.

Quite generally, wetting transitions at temperatures above the triple point
are expected for weakly attractive substrates. In particular, it was argued
that alkali metals provide the weakest adsorption potentials of any surfaces
for He atoms and therefore that wetting transitions could be observed
on such
substrates \cite{cheng1}. This hypothesis has subsequently been confirmed
and the corresponding transition studied in many laboratories \cite
{nacher,rutledge,ketola}. More recently, similar phenomena have also been
predicted and/or seen for H$_{2}$, Ne, and Hg on various surfaces \cite
{cheng2,ross1,mistura1,hess,hensel,kozhevnikov}.

Theory dictates \cite{nakanishi,dietrich} that when a first order transition 
from partial to complete wetting
occurs at a temperature $T_{w}$ above the triple-point (and below the bulk
critical temperature), then a locus of first-order surface phase transitions 
must extend away from the vapor-fluid coexistence curve, on the
vapour side. At $T<T_{w}$ the thickness of the adsorbed liquid film
increases continuously with pressure,\ but the film remains microscopically
thin up to the coexistence pressure $P_{sat}(T)$, and becomes infinitely
(macroscopically) thick just above it. The adsorption isotherms reach $%
P_{sat}(T)$ with a finite slope. At temperatures $T_{w}<T<T_{c}^{pw}$ ($%
T_{c}^{pw}$ being the prewetting critical temperature) the thin film grows
as the pressure is increased until a transition pressure $%
P_{tr}(T)<P_{sat}(T)$ is reached. At this pressure a thin film is in
equilibrium with a thicker one, and a jump in coverage occurs as $P$
increases through $P_{tr}(T)$. At still higher pressures this first order
transition is followed by a continuous growth of the thick film which
becomes infinitely thick at $P_{sat}(T)$. The discontinuity (i.e. the
difference in thickness between the two films in equilibrium) becomes
smaller and smaller as $T$ increases above $T_{w}$, until at a temperature $%
T_{c}^{pw}$ the two films are no longer distinguishable from one another.
Finally, when $T>T_{c}^{pw},$  the adsorbed liquid film
increases continuously with pressure to become infinitely thick at $P_{sat}(T)$.

Noble-gas fluids (other than He) adsorbed on weakly attractive substrates
such as the surface of alkali metals, are good candidates for showing
prewetting transitions. There is however a limited experimental knowledge
for these systems. The measurements of Hess et al.\cite{hess} indicate that
Ne undergoes a wetting transition at a temperature within $2\%$ of $T_{c}$
on a Rb surface and that it undergoes a drying transition on Cs. Non-wetting
of Ar on a Cs surface has been proposed on the basis of desorption
experiments \cite{friess}. No experimental information is available at
present on the wetting properties of Ar and Ne on other alkali surfaces.

In the absence of experimental observations, numerical simulations may play
an important role in understanding the wetting properties of fluids and may
be used as a useful guide in the choice of systems to be studied
experimentally. We present in this paper one such calculation, based on a
Density Functional approach, where the adsorption properties of simple
classical fluids (namely Ar and Ne) on Li and Na surfaces are studied.
Ne on Rb has also been studied as a test case, where the wetting 
transition has experimentally been observed \cite{hess}.
Recently, results of extensive Monte Carlo (MC) simulations on the Ne/Li
system have been presented\cite{bojan}, which seem to show that this system
is nonwetting for all T between the triple point and the critical point. As
shown in the following, our calculations predict quite a different scenario,
and are in fact consistent with prewetting transitions below the critical
point.
We will discuss the discrepancies between our findings and the MC
results of Ref.\cite{bojan} in Section 3.

\section{Computational Method}

In recent years Density Functional (DF) methods have become increasingly
popular because of their ability to describe inhomogeneous fluids and phase
equilibria. Comparison with simulation results shows that, once the
long-range (van der Waals) attractive forces exerted by a surface on the gas
atoms are included in the free energy density functional, the method
provides a qualitatively (and most often quantitatively) good description of
the thermodinamics of gas adsorption on a solid surface and in particular it
is able to predict correctly a large variety of phase transitions (wetting,
prewetting,layering,etc.).

In addition, also the microscopic structure of the fluid in the vicinity of
a solid surface, e.g. the oscillatory behavior of the density profile due to
''packing'' effects, may described accurately by the more sophisticated
Density Functionals, which treat in a non-local way the short-ranged
repulsive part of the intermolecular potential 
\cite{tarazona_all,meister,sokolowsky,vanderlick,kroll}.
In our calculations we use the DF
proposed in Ref.\cite{rosinberg} which has proven to describe successfully
the structure of a classical liquid in very demanding situations, e.g. in
the presence of hard-walls or strongly attractive surfaces, where rapid
variations of the liquid density occur in the close vicinity of the
adsorbing surface. Although this functional does not seem to describe
correctly inhomogeneities of infinite spatial extent such as those occurring
in a freezing transition, it is fairly accurate in describing density
inhomogeneities of finite extent.

In the Density Functional approach the free energy of the fluid is written
in terms of the density $\rho (\vec{r})$ of the fluid as: 
\begin{eqnarray}
F[\rho ] &=&F_{HS}[\rho ]+{\frac{1}{2}}\int \int \rho (\vec{r})\rho (\vec{r}%
^{\,\prime })u_{a}(|\vec{r}-\vec{r}^{\,\prime }|)d\vec{r}d\vec{r}^{\,\prime }
\nonumber \\
&&+\int \rho (\vec{r})V_{s}(\vec{r})d\vec{r}.  \label{energy}
\end{eqnarray}

Here $F_{HS}$ is the free-energy functional for an inhomogeneous hard-sphere
reference system, the second term is the usual mean-field approximation for
the attractive part of the fluid-fluid intermolecular potential $u_{a}$
\cite{allen},
while $V_{s}(\vec{r})$ is the external adsorption potential due to the
surface. For $F_{HS}$ we use the non-local functional of Ref. 
\onlinecite
{rosinberg} written in terms of a suitable coarse-grained density obtained
by averaging the true fluid density over an appropriate local volume. This
scheme predicts good triplet correlation functions $c^{(3)}({\bf r},{\bf r}%
^{\prime })$ for a bulk one-component fluid. In addition, applied to
liquid-solid interfaces it gives fairly good results, when compared to
''exact'' Monte-Carlo computer simulations, if the nonuniformities are not
too large \cite{rosinberg1}. For instance, both DF and MC simulations agree
in predicting and locating the prewetting transitions in the case of Ar/CO$%
_{2}$ system\cite{monson}. As an even more stringent test, the two-dimensional
limit of the theory of Ref.\cite{rosinberg} has been studied \cite{kr1}: the
behavior of a 3D DF theory in this limit is a good test of its performances
for describing adsorption phenomena at low coverages. The overall agreement
is surprisingly accurate \cite{kr1}.

One major weakness of the free-energy functional described above is its
mean-field treatment (i.e. its neglect of correlation effects) of the
long-range attractive part of the adatom-adatom interaction appearing in the
second term of Eq.(\ref{energy}).

This term gives the correct non-local dependence of the free energy with
respect to variations of the density at two points ${\bf r}$,${\bf r}%
^{\prime }$ away from each other, but it neglects the correlation effects
which should enhance the attractive interactions when $|{\bf r}-{\bf r}%
^{\prime }|$ is close to the minimum of the adatom-adatom potential. If one
includes this effect within perturbation theory, an expression for the free
energy similar to Eq. (\ref{energy}) should be used, but with the pair
correlation function of the adsorbed phase multiplying the interaction
in the integral of the second term.
Rather than
following this approach, which is cumbersome and impractical from a
computational point of view, we follow Ref. \onlinecite{tarazona} and
continue to use the much simpler form (\ref{energy}) but regard $u_{a}$ as
an {\it effective} attractive interaction which incorporates the main effect
of the pair correlation. This way of proceeding is found to improve the
prediction of the DF theory in many cases \cite{monson,tarazona}. We follow
previous prescriptions \cite{monson,tarazona,bruno} and  use in the second
term in Eq.(\ref{energy}) an effective interaction:

\begin{eqnarray}
u_{a}(r) &=&0\,\,\,\,\,,\,\,\,r\leq \lambda ^{\frac{1}{6}}\tilde{\sigma} 
\nonumber \\
&=&4\epsilon \{\lambda (\tilde{\sigma}/r)^{12}-(\tilde{\sigma}%
/r)^{6}\}\,\,\,\,\,\,,\,\,\,r>\lambda ^{\frac{1}{6}} \tilde{\sigma}
\label{pot}
\end{eqnarray}

which for $\lambda =1$, corresponds to the bare attractive interaction. For $%
\lambda <1$ this effective potential has the same long range properties as
the bare interaction, which is an important factor when studying wetting
phenomena, but its value at the minimum is increased by a factor $\lambda
^{-1}$, to simulate qualitatively the pair correlation effects described
above. At any temperature, we treat the HS diameter $\tilde{\sigma}$ \ and
the enhancing factor $\lambda $ as free parameters determined by requiring
that the experimental values of the liquid and vapor densities at
coexistence, $\rho _{l}$ and $\rho _{v}$, are reproduced for the bulk fluid,
as explained in detail in the following Section.

As for the interaction with the substrate, we make the usual approximation
of treating it as an inert, planar surface acting on the the fluid as an
external potential $V_{s}(z)$ ($z$ is the coordinate normal to the surface
plane). Accurate {\it ab initio} potentials are now available\cite{chizmeshya},
which describe the interaction between noble gases and the alkali surfaces.
Approximating the true surface by an ideal plane should not be a very bad
approximation for these systems, given the uniformity and almost
complete lack of corrugation of clean alkali surfaces.
The equilibrium density profile $\rho (z)$ of the fluid adsorbed on the
surface is determined by direct minimization of the functional in
Eq.  (\ref{energy})
with respect to density variations. In practice, at a given temperature,
we fix the coverage $\Gamma =\int (\rho (z)-\rho _{v})dz$ and solve
iteratively the Euler equation $\mu =\delta F/\delta \rho (z)$ by using a
fictitious dynamics, with the value of the chemical potential $\mu $ fixed
by $\Gamma $. The results of such minimizations are reported and discussed
in the following Section.

\section{Results and Discussion}

Our results are based on the analysis of adsorption isotherms, calculated by
using the DF described in the previous Section, for the case of Ar and Ne
adsorption on planar surfaces representing Li and Na surfaces, respectively.

The bare Ne-Ne and Ar-Ar interactions are usually given in the form of a
Lennard-Jones (LJ) 12-6 potential, with parameters \cite{hirsch,bojan}
$\epsilon _{Ne-Ne}/k=33.9\,\,K$, $%
\sigma _{Ne-Ne}=2.78\,\AA $ and $\epsilon _{Ar-Ar}/k=119.8\,\,K$, $\sigma
_{Ar-Ar}= 3.405\,\AA $.

We use in Eq.(\ref{pot}) 
the above values for $\epsilon $, while we determine the two adjustable
parameters entering the functional (\ref{energy}), i.e. the HS diameter $%
\tilde{\sigma}$ and the corrective factor to the potential well depth $%
\lambda $, by imposing that the experimental bulk phase diagram of the
adsorbed fluid (either Ar or Ne) is correctly reproduced. In particular, it
is crucial to verify that the choice of these two parameters leads to the
correct thermodynamic equilibrium conditions. We have calculated several
isotherms for a bulk system. If the temperature is lower than the critical
temperature then a blip in the $P-V$ plane, including region of positive
slope, $(\partial P/\partial V)_{T}>0$ , develops. As usual, the densities
of the coexisting liquid and gas and the equilibrium pressure are found by
applying a Maxwell (equal-area) construction in the P-V plane. For a
particular choice of $\tilde{\sigma}$ and $\lambda $, this construction
gives the vapor pressure $P$ and the liquid and vapor densities $\rho _{v}$
and $\rho _{l}$ at a given temperature. At any $T$, we determine the
parameters $\tilde{\sigma}$ and $\lambda $ giving the best fit to the
experimental \ values of $P(T)$, $\rho _{v}(T)$ and $\rho _{l}(T)$). 
A summary of our best fit parameters are given in Tables I and II, while
our results for the coexistence line in the $(\rho ^\ast ,P^\ast)$ plane
are shown in Fig. (\ref{fig1})
(we use reduced units for the 
quantities considered. These are defined as: $P^\ast
=P\sigma ^3/\epsilon $; $\rho ^\ast =\rho \sigma ^3 $, where $\epsilon $ 
and $ \sigma $ are the LJ parameters of the bare atom-atom interaction).

We notice that the value of $\lambda$ is such to increase the depth of the 
effective potential $u_a$ over the depth of the corresponding bare potential
by a factor $\simeq 1.78$ for Ar and $\simeq 1.8$ for Ne.
These values are of the same order of magnitude as that obtained in Ref.
\cite{bruno}  by fitting $T_c$ for Ar and are reasonable 
in order to describe the effect of the peak in $g(r)$ at the intermediate 
densities of interest. 
A second comment refers to the fitted values of the 
$\tilde \sigma$'s, which turn out to be about $5 \%$ smaller than the 
LJ values quoted above, usually employed in Monte-Carlo calculations 
\cite{bojan,bruno}.
As a consequence, the coefficients of the long range attraction, 
$C_6=4 \epsilon {\tilde\sigma}^6$ are reduced by about $30 \%$ with
respect to the corresponding above mentioned LJ values, and are in a much
better agreement with the most accurate theoretical values \cite{tang}.  
As remarked in Ref.\cite{allen}, however, 
the LJ parameters have nothing of fundamentals and are only determined to
provide reasonable agreement between "exact" results (say, from Montecarlo simulations) 
and the
experimental results in the bulk liquids. In other words they are not 
the values
which would apply to an isolated pair of Ar or Ne atoms. 
Similarly, our $\tilde\sigma$'s are fitted so that the values predicted 
by our DF calculation for the coexistence pressure at a given T and for 
the corresponding
densities agree with their experimental counterparts. In view of this, our 
effective interaction entering Eq.(\ref{pot}) 
should not be regarded as a pair potential and we do not
expect that,  when inserted in a Montecarlo simulation, it would necessarily
lead to the same results as those found from our DF treatment.
 
Once the bulk properties are optimized in the way described above, 
we switch on the adatom-surface
potential in Eq. (\ref{energy}) and compute the chemical potential $\mu $
corresponding to the equilibrium density profile for a given coverage, as
described in Section 2. The coverage is expressed in nominal layers $l=\rho
_{l}^{-2/3}\int_{0}^{\infty }[\rho (z)-\rho _{v}]dz$. The collection of
pairs ($\mu ,l$) at equilibrium, at a given $T$, represents an adsorption
isotherm.
As an example the isotherm of Ar/Li at $T=128 K$ is shown in Fig.(\ref{fig2}),
where the chemical potential, measured with respect to its value 
at coexistence, is plotted as
a function of the film thickness . For small coverages the chemical potential
increases approaching the saturation value from below. For certain
temperatures like the one considered in Fig.(\ref{fig2}), however, a sudden
change in slope is observed as the coverage is further increased. After
reaching a minimum value the chemical potential rises again, generating in
the $\mu -l$ plane a van der Waals-like loop. 
Since films for which $\mu $ has a negative slope are unstable,
this structure reveals the existence of first-order transitions between
films of different thickness. The amplitude of the transition (i.e. the
amplitude of the discontinuous jump in coverage) is determined by making an
equal-area Maxwell construction. The results of such a construction are
shown in Fig.(\ref{fig2}) with a dashed line: two equilibrium thicknesses
are thus identified, $l_{1}$ and $l_{2}$, and the jump between them
represents a prewetting transition.

A few equilibrium density profiles are shown in Fig.(\ref{fig3}), for
coverages smaller and larger, respectively, than the thin-thick film values $%
l_{1}$ and $l_{2}$ determined above.

We summarize our results for the Ar/Li system in Fig.(\ref{fig4}), where the
calculated adsorption isotherms for this system are shown for a number of
temperatures. It appears from Fig.(\ref{fig4}) that a wetting transition 
occurs at $T\simeq 124\,K$, accompanied by  first-order prewetting
transitions at higher $T$.  The amplitude of the prewetting transition
decreases with temperature, until it vanishes at a critical value $%
T_{c}^{pw}\simeq 130\,K$. At temperatures higher than $T_{pw}^{c}$ a
continuous wetting of the surface takes place.

In the $(\mu - T)$ plane the occurrence of these first order wetting transitions
is represented by a line leaving smoothly
the liquid-gas coexistence curve at $T_w$ and
ending at the prewetting critical temperature $%
T^c_{pw}$.
On general grounds one expects $\Delta \mu \equiv \mu - \mu _{coex} \sim
-a(T-T_w)^{3/2} $, where the exponent $3/2$ is related to the exponent of
the van der Waals tail $-C/z^3$ of the surface potential. We show in Fig.(%
\ref{fig5}) our calculated values for $\Delta \mu $, together with a fit
with the expected analytical form.

Fig.(\ref{fig6}),(\ref{fig7}),(\ref{fig8}) show the calculated adsorption
isotherms for  Ar/Na, Ne/Li and Ne/Na, respectively. In all cases a
sequence of prewetting transitions is found, below the bulk critical
temperature ($T_c\,(Ar)=150.9 K$ and $T_c\,(Ne)=44.4 K$).
In Table III we summarize the wetting and critical pre-wetting temperatures
obtained from our calculated adsorption isotherms. In parenthesis we
report the wetting temperature as predicted by Eq.(\ref{criterion}) (see the
following).

The last line in Table III refers to the system Ne/Rb
experimentally investigated by Hess et al.
\cite{hess}. 
We calculated two isotherms for this system, at $T=43\,K$ and $T=44\,K$.
We find non-wetting behavior at $T=43\,K$, while at $T=44\,K$ complete wetting
occurs. We thus estimate $T_w\sim 44\,K$ for this 
system, but we are not able to resolve any prewetting line of
transitions between these two temperatures.
Both the wetting temperature and the temperature interval $|T_w-T_{c}^{pw}|\sim 1\,K$ 
separating the
nonwetting from the complete continuous wetting regimes 
are in excellent agreement with experiments \cite{hess}.
Although the almost perfect agreement between theoretical and 
experimental values of $T_w$ should
be regarded as fortuitous in view of the many approximations contained in our
calculations ( planar surface, DF treatment, effective potential fit etc.),
nonetheless it is rewarding to find that our treatment correctly predicts the
observed wetting transition. 

A simple  heuristic model has been proposed by
Cheng et al. \cite{cheng} where the energy cost of forming a thick film is
compared with the benefit due to the gas-surface attractive interaction,
resulting in an implicit equation for the wetting temperature 
\begin{eqnarray}
(\rho _l-\rho _v)\int_{z_0}^\infty V_s(z)dz=-2\gamma  \label{criterion}
\end{eqnarray}

Here $\rho _{l}$ and $\rho _{v}$ are the densities of the adsorbate liquid
and vapor at coexistence, $\gamma $ is the surface tension of the liquid and 
$z_{0}$ is the equilibrium distance of the 
gas-surface interaction potential $V_{s}$. If one uses in
Eq.(\ref{criterion}) the potential $V_{s}$ used in our calculations, and the
experimental values for $\rho _{l}(T)$, $\rho _{v}(T)$ and $\gamma (T)$,
values of the wetting temperature, $T_{w}$ systematically lower than those
found in our calculations are predicted, as shown in the column $T_{w}^{C}$
of Table III. One issue raised by the present results is thus an
apparent failure of this simple model to describe wetting behavior in the case
of ultraweak potentials, such as those investigated here.

Prewetting transitions in classical fluids are notoriously elusive.
On the basis of model calculations, 
it has been suggested \cite{sen} that
one possible reason for this elusiveness is that they lie so close to the
adsorbate bulk coexistence line as to render them difficult to detect. This
happens because the adatom-substrate potential is comparable in strength
with the adatom-adatom potential. The model calculations of Ref.\cite{sen}
show that, for a wide range of interaction parameters, the prewetting line
lies at a chemical potential which differs from that of coexistence
by an amount on the order of $10^{-3}KT_{c}$. Our results show  no exception
to this behavior, although the proximity to coexistence, which makes
prewetting transitions so difficult to observe experimentally, is slightly
lower than expected. In the case of Ne/Li the values of the chemical
potential for which the discontinuities in  coverage are observed are only $1K
$ (at most) below the saturation value (see Fig. 7), while for Ar/Li 
they are $2K$ (at most) below (see Fig. 4). 
Thus $\Delta \mu \sim 10^{-2}KT_{c}$. 
The vapor
pressure at which the prewetting jumps should be observed for the system we
have investigated, may be approximately estimated 
 by assuming  ideal gas behavior for the
vapor phase (see Ref. \cite{bojan} for a better estimate).
One finds $P=P_{sat}exp(\Delta \mu /T)\sim 0.985P_{sat}$ for $%
\Delta \mu \sim 2\,K$, thus in a range which may be 
accessible to experiments.

We mention at this point recent results\cite{bojan} based on extensive Monte
Carlo simulations on the behavior of \ Ne on a Li surface (the gas-surface
potential used in these simulations is the same as used here ) where
nonwetting behavior up to the critical point has been found, at variance
with the results presented here. One may invoke various explanations for
these different findings. For example one could think that metastability in
the region close to the critical point may affect seriously MC simulations
and that reliable results may require a number of MC moves prohibitively
large to reach the  regime where the wetting transition is found. 

Another possible explanation on why Boyan et al. do not see wetting
of Ne on Li is that the prewetting 
regime is
so close to saturation that their simulation cannot discern it. 
In fact at 40 K we find a prewetting jump of slightly more than two layers at a
chemical potential only $0.5 K$ below saturation. At higher temperatures, 
where the jump is higher,  it occurs even closer to saturation. 
The situation is even worse for Ne/Na. It seems to be more promising 
for Ar/Li, where one should see a $\sim 2-3$ layers jump $\sim 1.5 K$ 
below saturation at 
$T=128 K$, so this is the system were we
suggest to make experiments.

We remark
at this point the extreme sensitivity of the results presented above to the
details of the gas-surface interaction potential. It has been shown recently\cite
{ancilotto}, for the case of Ar/CO$_{2}$ system, that minor changes
in the overall shape of the adsorption potential may alter dramatically the
properties of the adsorbed film. In order to evaluate the sensitivity of the
present results to the shape of the substrate potential,
we re-calculated the wetting
diagram of Ar/Li by using a 9-3 Lennard-Jones potential to describe
the Ar-surface interaction, instead than
the ab-initio potentials used above \cite{chizmeshya}, 
and adjusting the two parameters entering the 9-3 potential
in order to have the same well depth $D$ and minimum position as the 
ab-initio potential.
We found that the wetting temperature 
calculated with this interaction shifts upward from 123 to 136 K, while the
critical wetting temperature changes from 130 to 138. This is not
surprising, given the dependence of the wetting temperature, as clear from
the implicit definition Eq. (\ref{criterion}), from the global shape of the 
potential, and not
just from the well depth and position of the minimum. 

Of course the wetting properties will also be very sensitive to the
adatom-adatom potential, which determines the surface tension explicitly
entering Eq. (\ref{criterion}). Thus, on these very weak substrates,
wetting or non wetting above the triple point, but below $T_c$, is the result
of a delicate balance between adatom-substrate and adatom-adatom interactions.
Another possible explanation of the discrepancy between our findings and
the MC results of Bojan et al. \cite{bojan} may thus be ascribed to the
different long range behavior of our effective interaction $u_a$ (see Eq.(2) 
and the ensuing discussion) and
of the bare LJ potential used in the Montecarlo simulations of Ref.\cite{bojan}.

Experiments are currently in progress
\cite{mist1}, which will hopefully be able to verify the predictions
contained in this paper. 

\acknowledgments
We thank G.Mistura, L.Bruschi, M.W.Cole and S.Curtarolo 
for useful discussions and critical
comments.

\bigskip

\begin{figure}[htb]
\caption{ Phase diagram of bulk-Ar. Squares: calculated points. Solid line:
experimental curve.}
\label{fig1}
\end{figure}

\begin{figure}[htb]
\caption{ Chemical potential (measured with respect to its value at bulk
liquid-vapor coexistence) as a function of coverage for the Ar/Li system at 
T=128 K. The dashed line shows the value of $\Delta \protect\mu $ satisfying
the equal-area Maxwell construction: the thickness of the "thin" and "thick"
film in equilibrium at this value are indicated on the x-axis. }
\label{fig2}
\end{figure}

\begin{figure}[htb]
\caption{ Density profiles of Ar films on the Li surface, plotted as a function
of the distance from the surface plane.
The thin lines
show the density profiles at coverages below the "thin" film thickness $l_1$
(see Fig.2), the thicker lines show the density for film thickness larger
than the value $l_2$ shown in Fig.2. 
The origin of the $z$ coordinates is
taken at the surface plane position.}
\label{fig3}
\end{figure}

\begin{figure}[htb]
\caption{ Adsorption isotherms for the Ar/Li system. $\Delta \protect\mu $
is measured from the saturation value. Squares: $T=124\,K$; Full dots: $%
T=126\,K$; Triangles: $T=128,K$; Crosses: $T=129\,K$; Stars: $T=130\,K$;
Open dots: $T=132\,K$. }
\label{fig4}
\end{figure}

\begin{figure}[htb]
\caption{ Calculated wetting phase diagram for the Ar/Li system. The squares
are the calculated points, the dashed line is a fit, as described in the
text. }
\label{fig5}
\end{figure}

\begin{figure}[htb]
\caption{ Adsorption isotherms for the Ar/Na system. Squares: $T=137\,K$;
Full dots: $T=138\,K$; Triangles: $T=139,K$; Crosses: $T=140\,K$. }
\label{fig6}
\end{figure}

\begin{figure}[htb]
\caption{ Adsorption isotherms for the Ne/Li system. Squares: $T=39\,K$;
Full dots: $T=40\,K$; Triangles: $T=41\,K$. }
\label{fig7}
\end{figure}

\begin{figure}[htb]
\caption{ Adsorption isotherms for the Ne/Na system. Squares: $T=41\,K$;
Full dots: $T=42\,K$; Triangles: $T=43\,K$. }
\label{fig8}
\end{figure}

\bigskip
\bigskip

\begin{table}[tbp]
\caption{ Values of $\tilde\protect\sigma  $ and $\protect\lambda $ used to
obtain the agreement between the experimental and calculated values of
coexistence densities for bulk-Ar at
the saturation pressure, as explained in the text. Calculated
values are reported in parenthesis. }
\label{table:param1}
\begin{tabular}{cccccc}
$T \,( K ) $ & $P^\ast $ & $\rho _v ^\ast $ & $\rho _l ^\ast $ & $\tilde%
\sigma/\sigma_{Ar-Ar}$ & $\lambda $ \\ \hline
124 & 0.036 (0.037) & 0.044 (0.044) & 0.669 (0.669) & 0.9562 & 0.5810 \\ 
126 & 0.040 (0.041) & 0.049 (0.048) & 0.658 (0.658) & 0.9545 & 0.5815 \\ 
128 & 0.044 (0.046) & 0.055 (0.054) & 0.646 (0.646) & 0.9520 & 0.5840 \\ 
129 & 0.046 (0.047) & 0.058 (0.057) & 0.638 (0.638) & 0.9526 & 0.5838 \\ 
130 & 0.048 (0.050) & 0.061 (0.060) & 0.634 (0.634) & 0.9508 & 0.5833 \\ 
132 & 0.053 (0.055) & 0.068 (0.067) & 0.621 (0.621) & 0.9485 & 0.5855 \\ 
137 & 0.066 (0.070) & 0.089 (0.087) & 0.586 (0.586) & 0.9438 & 0.5886 \\ 
138 & 0.069 (0.073) & 0.094 (0.092) & 0.578 (0.578) & 0.9428 & 0.5897 \\ 
139 & 0.073 (0.077) & 0.100 (0.098) & 0.569 (0.569) & 0.9416 & 0.5912 \\ 
140 & 0.076 (0.081) & 0.106 (0.103) & 0.561 (0.560) & 0.9404 & 0.5923
\end{tabular}
\end{table}

\begin{table}[tbp]
\caption{ Same as in Table I, for bulk-Ne. }
\label{table:param2}
\begin{tabular}{cccccc}
$T \,( K ) $ & $P^\ast $ & $\rho _v ^\ast $ & $\rho _l ^\ast $ & $\tilde%
\sigma/\sigma_{Ne-Ne}$ & $\lambda $ \\ \hline
39 & 0.058 (0.061) & 0.074 (0.072) & 0.596 (0.596) & 0.9537 & 0.5526 \\ 
40 & 0.067 (0.071) & 0.088 (0.086) & 0.574 (0.574) & 0.9497 & 0.5536 \\ 
41 & 0.077 (0.083) & 0.106 (0.103) & 0.549 (0.549) & 0.9455 & 0.5549 \\ 
42 & 0.089 (0.096) & 0.127 (0.124) & 0.520 (0.520) & 0.9413 & 0.5567 \\ 
43 & 0.102 (0.104) & 0.156 (0.138) & 0.482 (0.482) & 0.9470 & 0.5503 \\
44 & 0.116 (0.123) & 0.206 (0.201) & 0.417 (0.416) & 0.9347 & 0.5695           
\end{tabular}
\end{table}

\begin{table}
\caption{Wetting ($T_w$) and critical prewetting ($T_{pw}^c $)
temperatures from DF calculations. The column labeled $T_w^{C}$ reports
the prediction of Eq. (\ref{criterion})}.
\begin{tabular} {c c r r r}
Gas / Surface & $T_w$ & $T_{pw}$ & $T_w^{C}$ & $T_w^{expt}$ \\ \hline
Ar/Li   & 123  (0.81$\,T_c$) &   130 & 107 & - \\
Ar/Na   & 136  (0.90$\,T_c$) &   140 & 117 & - \\
Ne/Li   & 38   (0.86$\,T_c$) &     41 & 33 & - \\
Ne/Na   & 40   (0.90$\,T_c$) &     44 & 36 & - \\
Ne/Rb   & 44   (0.99$\,T_c$) &     - & 38 & 43 \\
\end{tabular}
\label{table:param}
\end{table}

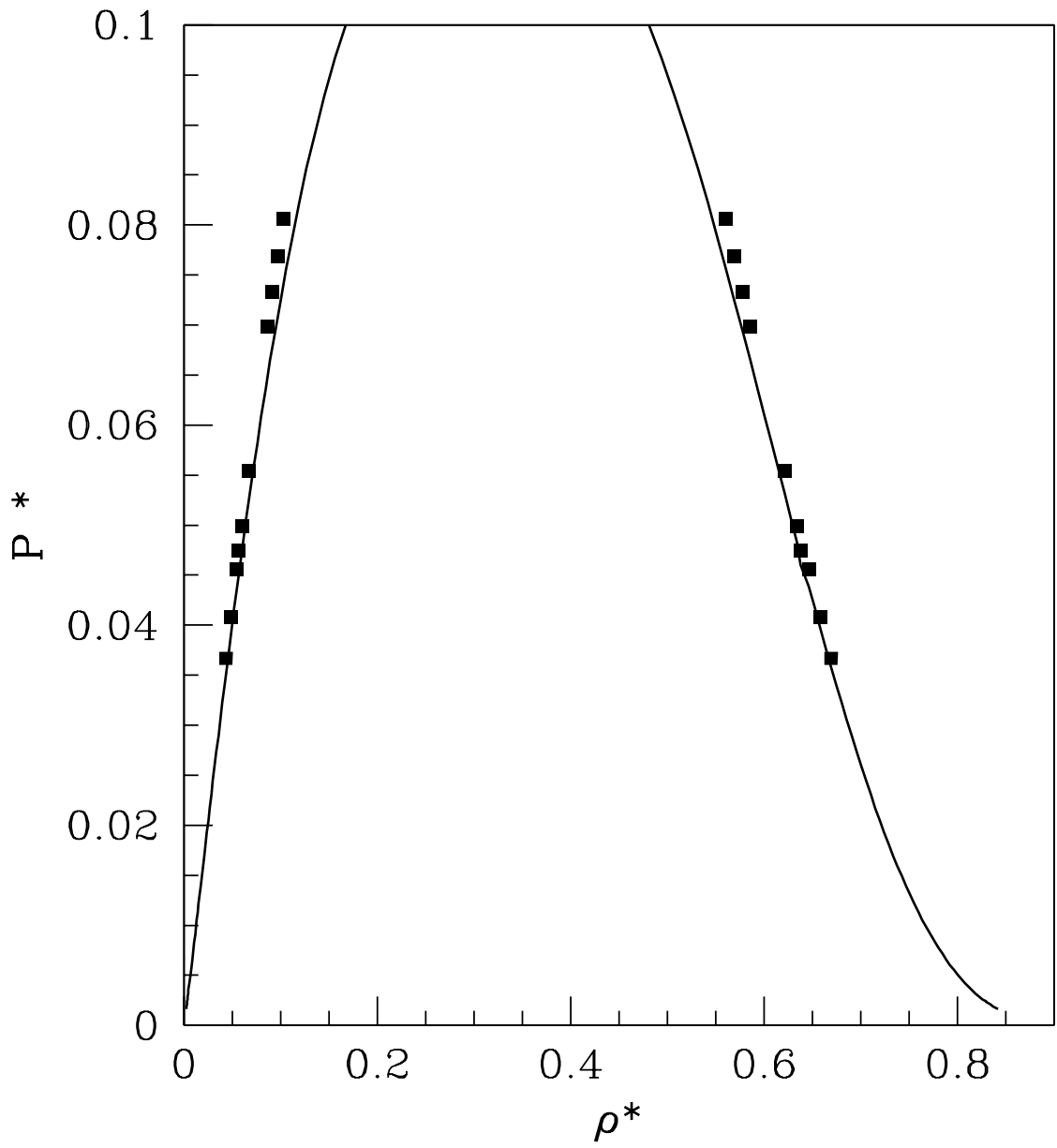
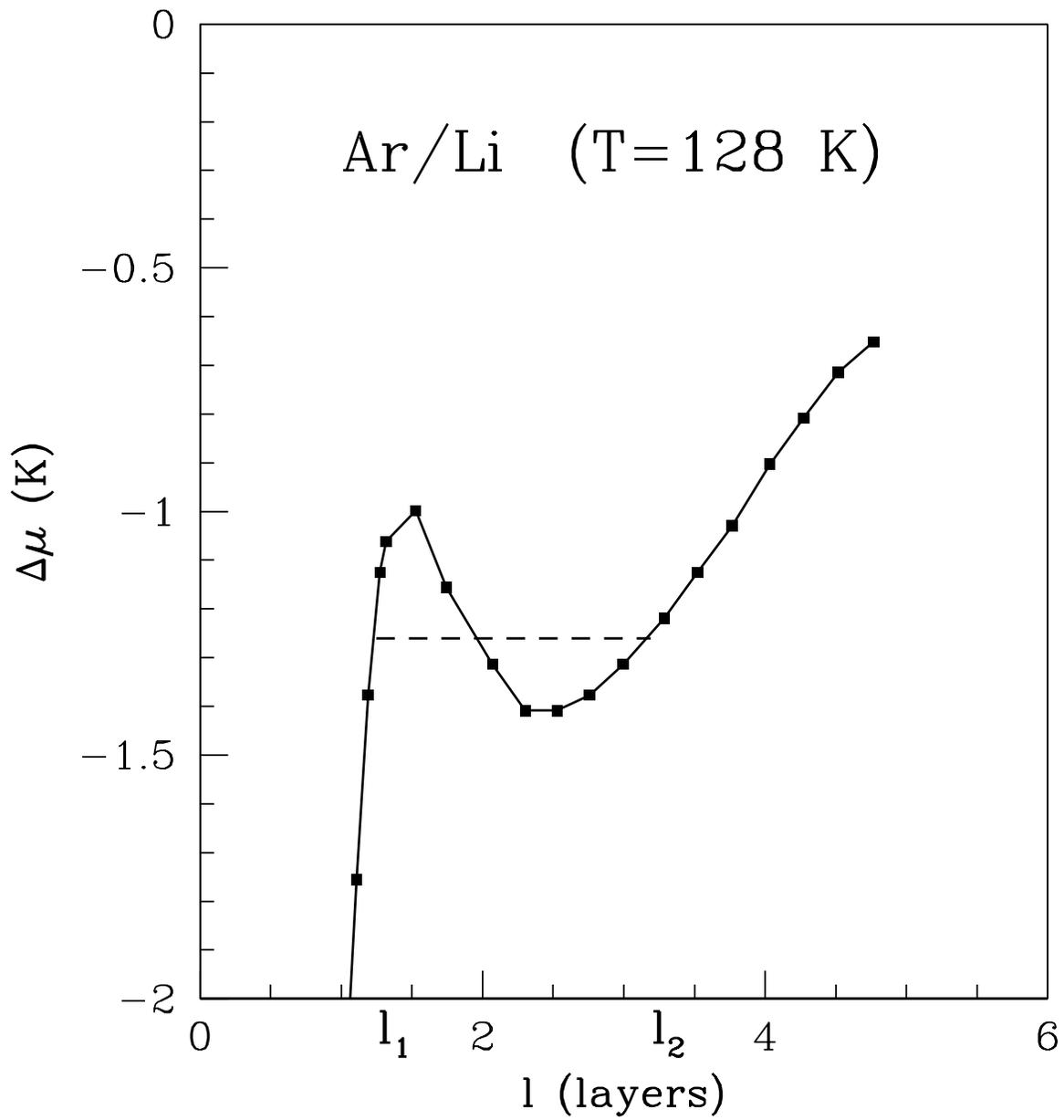
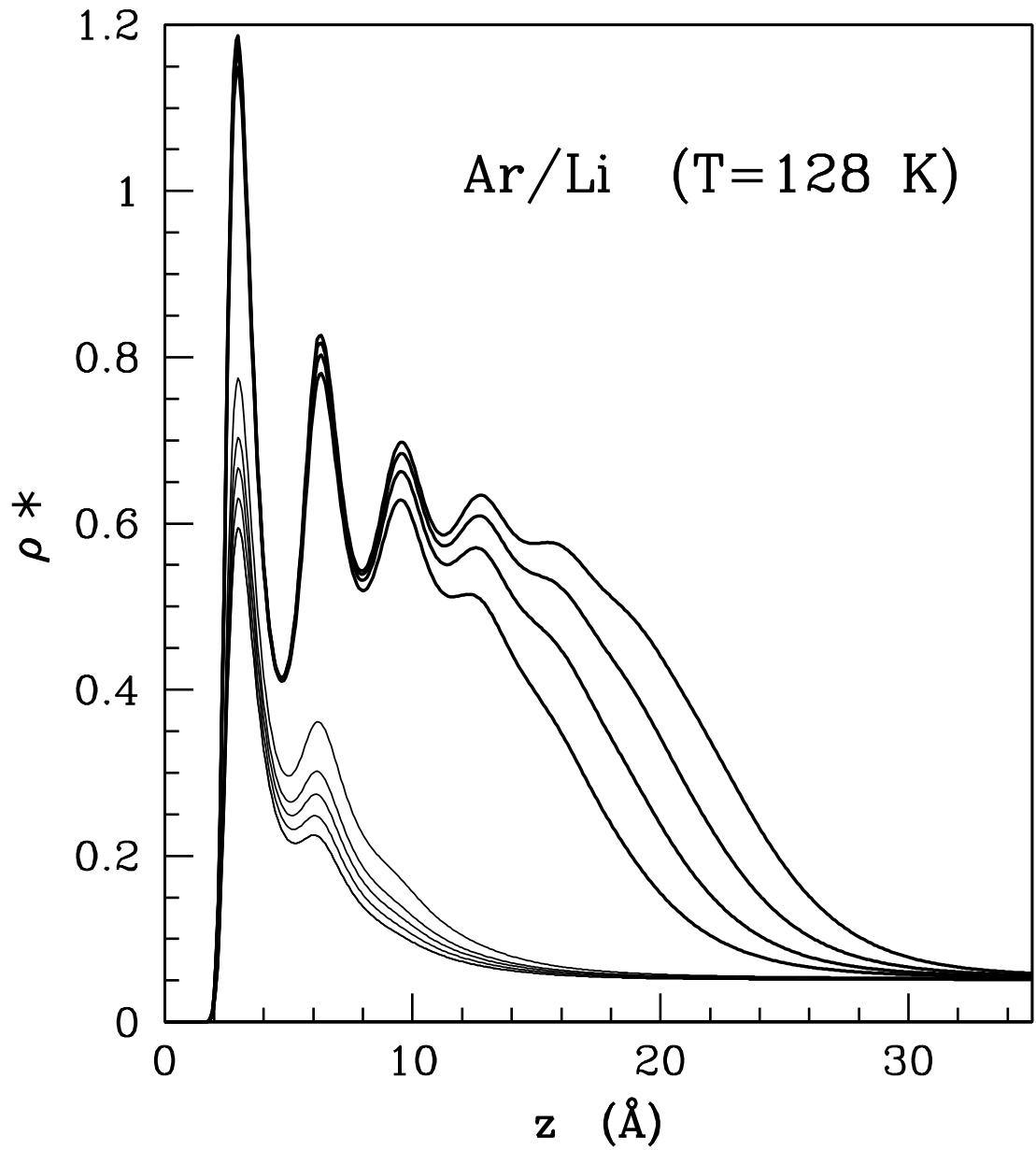
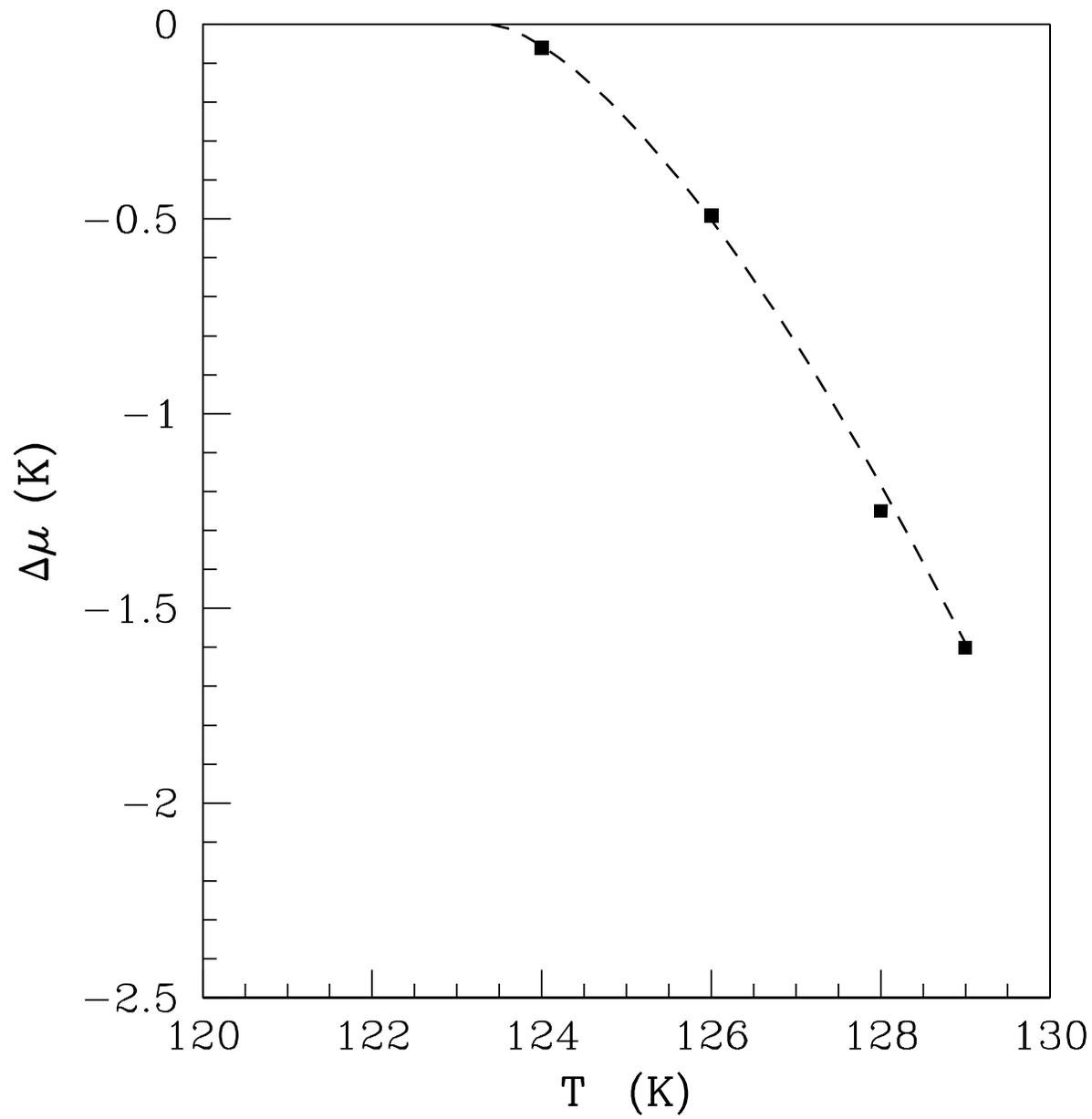
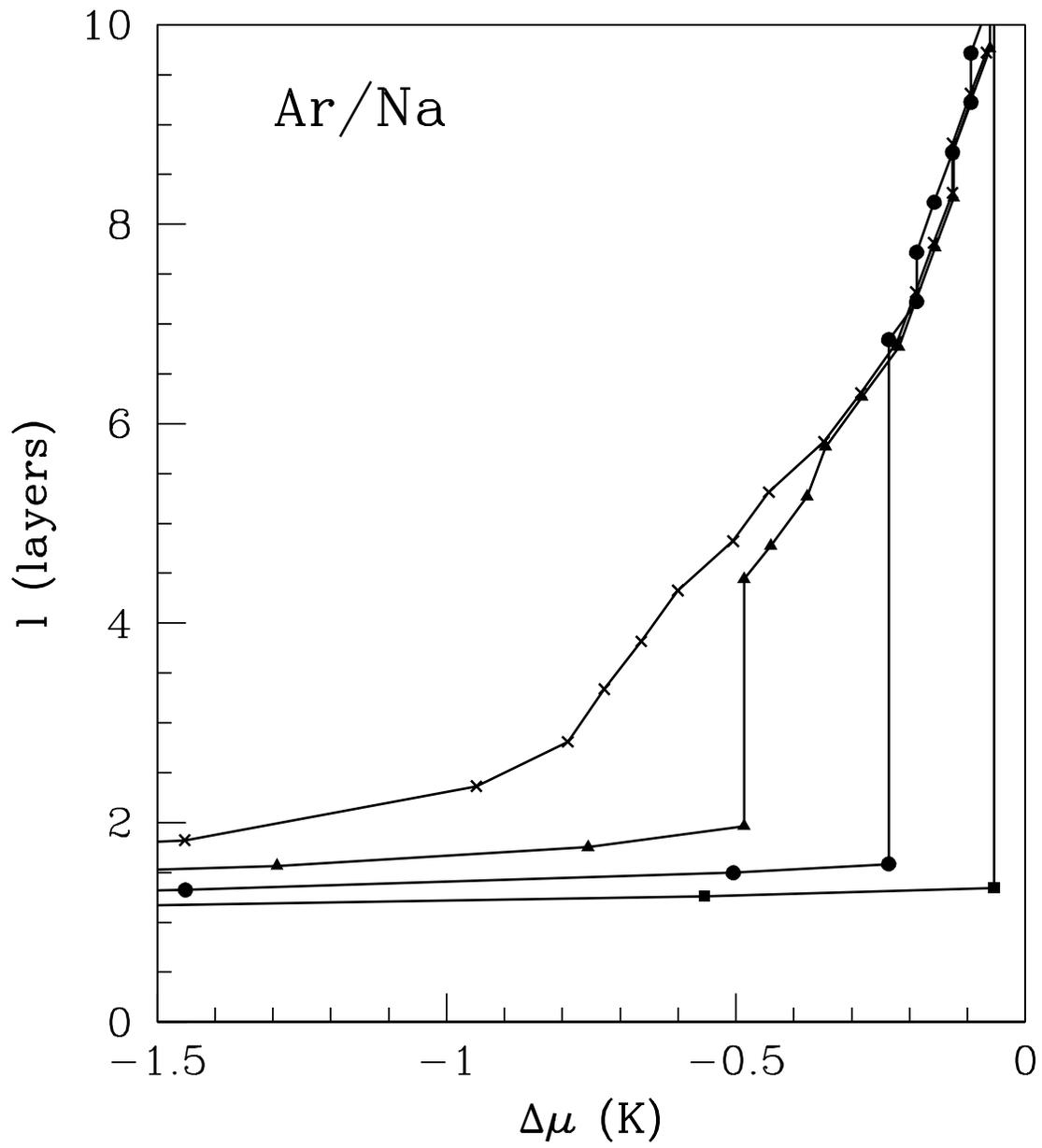
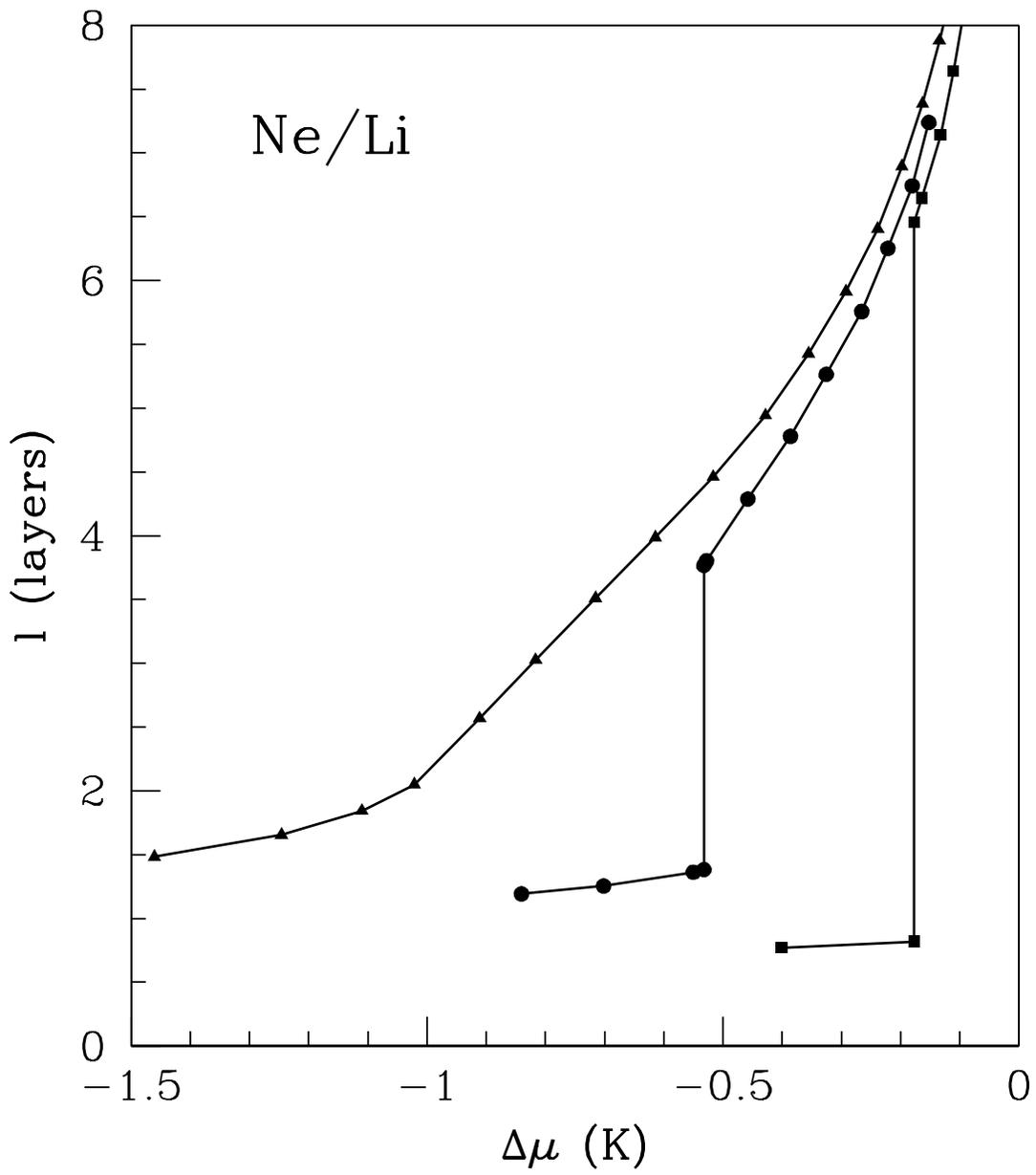
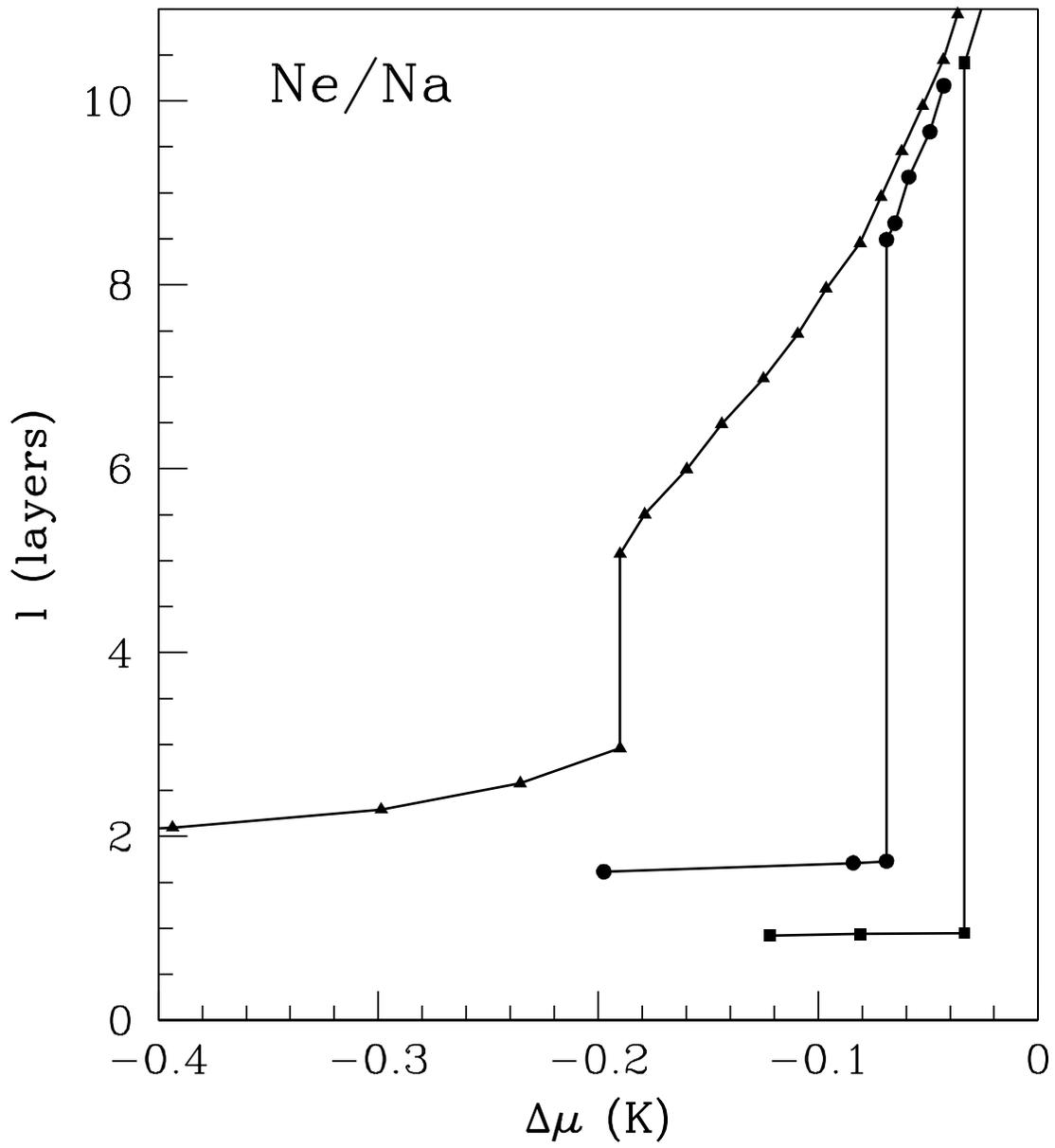
\end{document}